\documentclass[reprint,5p,times]{elsarticle}

\usepackage[utf8]{inputenc}
\usepackage{amsmath}
\usepackage{amssymb}
\usepackage{bbold}
\usepackage{amsbsy}
\usepackage{bm}
\usepackage{graphicx}
\usepackage{array}
\usepackage{tabularx}
\usepackage{mathtools, cuted}

\usepackage{calrsfs}
\DeclareMathAlphabet{\pazocal}{OMS}{zplm}{m}{n}

\newcommand{\vsig}{\boldsymbol{\sigma}}
\newcommand{\veps}{\boldsymbol{\epsilon}}
\newcommand{\vu}{\mathbf{u}}
\newcommand{\vz}{\mathbf{0}}
\newcommand{\pe}{\! = \!}
\newcommand{\cE}{\mathcal{E}}
\newcommand{\cH}{\mathcal{H}}
\newcommand{\bcE}{\bar{\cE}}
\newcommand{\fud}{\mbox{$\frac{1}{2}$}}
\newcommand{\xt}{\tilde{x}}
\newcommand{\hti}{\tilde{h}}
\newcommand{\hs}{\hti_s}
\newcommand{\hu}{\hti_1}
\newcommand{\Lt}{\tilde{L}}
\newcommand{\Ht}{\tilde{H}}
\renewcommand{\tt}{\tilde{t}}
\newcommand{\ls}{l^*}
\renewcommand{\le}{{\lambda^*}}
\newcommand{\ts}{\tau^*}
\newcommand{\ATG}{{\scriptscriptstyle{ATG}}}

\begin{document}

\begin{frontmatter}

\title{Shape relaxation of epitaxial mesa for finite-size strain-engineering}

\author[1]{Kennet D. R. {Hannikainen}}

\author[2]{Luc Favre}

\author[3]{Fabien Deprat}

\author[4]{Olivier Gourhant}

\author[5]{Isabelle Berbezier}

\author[6]{Jean-No\"el Aqua\corref{cor1}}
\ead{aqua@insp.jussieu.fr}

\cortext[cor1]{Corresponding author}

\affiliation[1]{organization={Sorbonne Université, Centre National de la Recherche Scientifique, Institut des NanoSciences de Paris, INSP}, addressline={4 place Jussieu}, postcode={F-75005}, city={Paris}, country={France}}

\affiliation[2]{organization={Aix Marseille Université, CNRS, Université de Toulon, IM2NP}, addressline={Avenue Escadrille Normandie Niemen}, postcode={13397}, city={Marseille}, country={France}}

\affiliation[3]{organization={ST-Microelectronics}, addressline={850 Rue Jean Monnet}, postcode={38920}, city={Crolles}, country={France}}

\affiliation[4]{organization={ST-Microelectronics}, addressline={850 Rue Jean Monnet}, postcode={38920}, city={Crolles}, country={France}}

\affiliation[5]{organization={Aix Marseille Université, CNRS, Université de Toulon, IM2NP}, addressline={Avenue Escadrille Normandie Niemen}, postcode={13397}, city={Marseille}, country={France}}

\affiliation[6]{organization={Sorbonne Université, Centre National de la Recherche Scientifique, Institut des NanoSciences de Paris, INSP}, addressline={4 place Jussieu}, postcode={F-75005}, city={Paris}, country={France}}




\date{\today}

\begin{abstract}
Silicon-Germanium (Si$_{1-x}$Ge$_x$) layers are commonly used as stressors in the gate of MOSFET devices. They are expected to introduce a beneficial stress in the drift and channel regions to enhance the electron mobility. When reducing the gate lateral size, one of the major issues is the stress relaxation which results in a significant decrease in the electron mobility. 
We report a new morphological evolution of a strained epitaxial SiGe nanolayer on a silicon gate (mesa) driven by strain inhomogeneity due to finite-size effects. Unlike the self-induced instability of strained films, this evolution arises here due to the elastic inhomogeneity originating from the free frontiers. We analyze the growth dynamics within the thermodynamic surface diffusion framework accounting for elasticity and capillarity, the former being solved in two dimensions thanks to the Airy formalism. The resulting dynamical equation is solved with a decomposition on eigenmodes, and reveals different developments depending upon the mesa geometric parameters. Mass transfer occurs towards the relaxed areas and creates a beading at the nanolayers free surface with either a W  or V shape as a function of time and geometry. The evolution is then controlled by the proportions of the structure as well as its scale. 
\end{abstract}




\end{frontmatter}


\section{Introduction}

The “Internet of Things” (IoT) which is growing at an amazing rate and is expected to continue to accelerate, should connect in the future years tens of billions of  physical devices embedded with tiny computing devices that can sense and communicate from anywhere coming online, in an effort to improve the communications, flexibility, and customization of our daily needs and activities. These achievements are based on the pursuit of the miniaturization associated with the reduction of the energy consumption that should be achievable thanks to unrivalled integration expected for the next generation Complementary Metal Oxide Semiconductor (CMOS) architectures. The CMOS transistor is the basic building block for logic devices which represents the digital content of an integrated circuit. With the transistor down-scaling, the performances, density and costs have progressed by several decades. In p-channel MOSFETS (p-MOSFETS) SiGe source and drain were used as stressors to strain the channel and increase hole mobility. SiGe is also used as a CMOS higher mobility channel material. Compared to III-V semiconductors, SiGe has the advantage of CMOS compatibility and more symmetric carrier mobilities.
The use of SiGe strained channel became prevalent at the 65 nm node and has continued to be a large part of logic device performance improvements in every technology generation. In addition, the integration of these strained layers will continue to be crucial for the fabrication of the channels of bipolar transistors (in BiCMOS55 and BiCMOS28) and beyond since all the ultimate CMOS technologies use selective epitaxy. In recent and future technology nodes starting from and beyond 28nm, epi SiGe layers are integrated in all MOSFETs and they are the central part or the active areas of the transistors: drain/source (D/S) recessed or raised, or channel material in order to increase the holes mobility. Epi SiGe are also used to tune the threshold voltage and to improve the access resistance of pMOS which can be roughly aligned with nMOS, making the CMOS technologies much more efficient. 
Since the morphology of the SiGe channels will impact the function of the final device, it is crucial to characterize their morphological evolution when confined in small areas during deposition as well as during post-deposition thermal budgets. In the same way, the biaxial strain in the MOS channel plays a major role on the electrical performances and has to be controlled in both directions, parallel and perpendicular to the D/S current. 

In this context, SiGe based devices are extensively studied for the end-of-roadmap CMOS for telecom, mobiles and set top boxes. The aim of this manuscript is to provide an extensive and fundamental understanding of the channel behavior on short gates and to predict the evolution for the next 28nm generation processes. It is part of all studies aimed at using the finite size effects to control the physical properties of research or industrial devices. For example, the development of the fully-depleted complementary metal oxide semiconductors (FD-CMOS) technology \cite{FeraColi11} is based on ultra-thin silicon-on-insulator (SOI) \cite{Brue95} for which strain engineering in Si and SiGe layers of gate and source/drain areas is a major technological challenge \cite{WebeJosse14,BaeBae16}. The unique features of nanoscale patterns can offer dramatic changes in growth modes as well as matter and strain redistribution that are conceptually new as compared to bulk substrates given that finite-size effects come significantly into play. Strain may be used to change the band structure, effective mass and mobility \cite{RiegVogl93,NayaChun94} and boost the device performance \cite{GhanArms03,ThomArms04}. The use of e.g.~SiGe intrinsically strained channels with a high level of strain is hence promising both for FinFET and FDSOI technologies. 

Of special interest for different technological development is the growth of epitaxial layers either on accommodating pseudo-substrates \cite{BerbAqua14,AquaFavr15} or on patterned substrate. Here, we are concerned by the growth of SiGe layers on top of a Si mesa obtained e.g.~with mesa etching. This enhances finite-size effects relative to bulk effects. Strain distribution in similar patterned films has been investigated, see e.g.~\cite{Hu91,Faux94,JainHark95}, and revealed the inhomogeneous relaxation that may have an impact on the subsequent electronic properties. However the influence of this patterning on the morphological evolution of the film has not been investigated yet, which is the focus of this article. We thence develop a dynamic model of the relaxation related to finite size effects, based on the analytical resolution of elasticity by the Airy function.


\section{Elastic fields}

\subsection{Airy function}

\begin{figure}[!ht] \centering
        \includegraphics[width=0.45\linewidth]{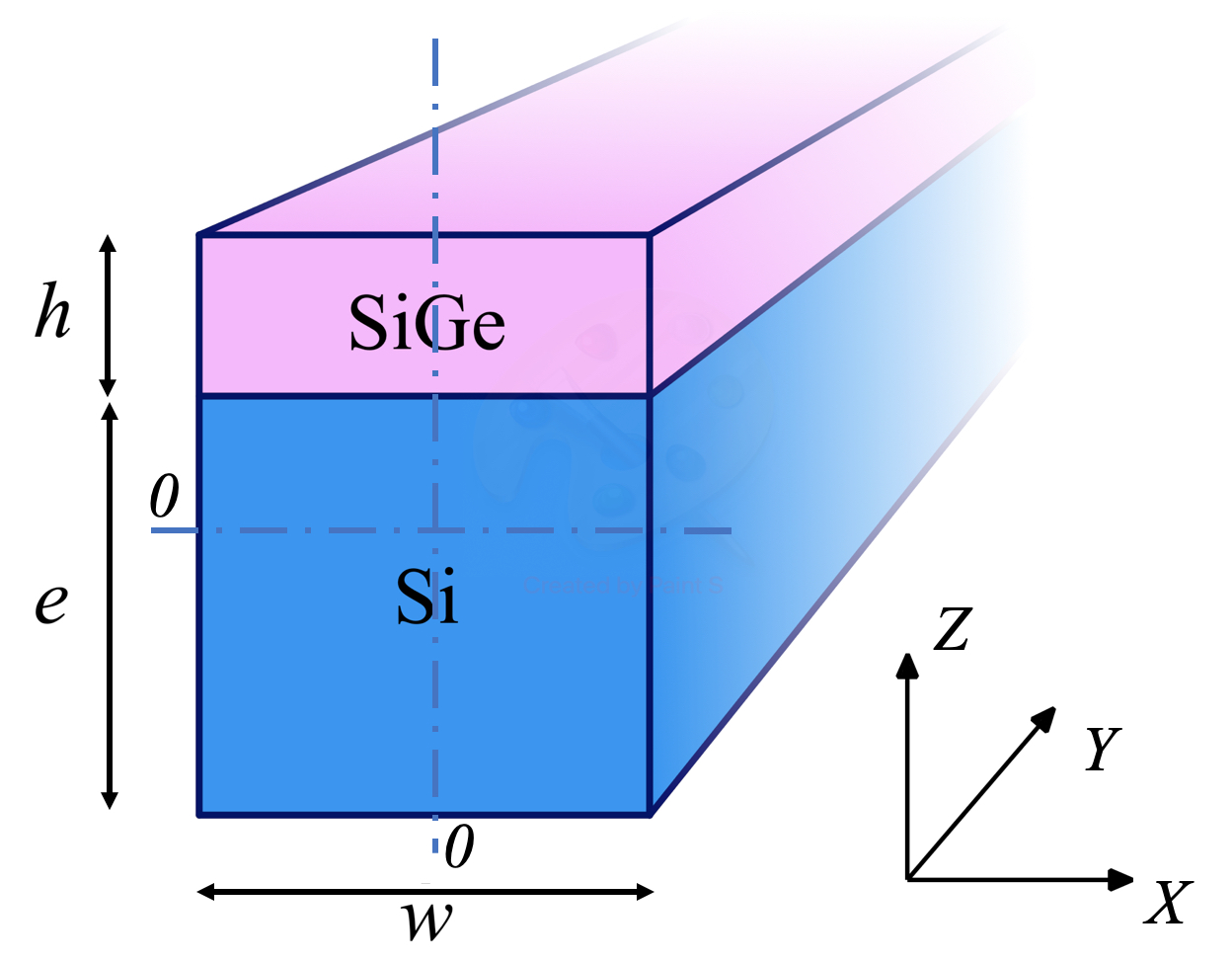}
    \caption{Schematic of the mesa system under consideration. A film of thickness $h$ is coherently deposited on a mesa of thickness $e$ and width $w$. The system is deposited on a semi-infinite substrate (not displayed). (color online)}
    \label{fig-geom}
\end{figure}

We consider a film of thickness $h$, epitaxially deposited on a mesa with a finite width $w$ and height $e$, see Fig.~\ref{fig-geom}. The film is bound to evolve due to surface diffusion during annealing at constant $h$ and temperature $T$. Given that mechanical equilibirum occurs under a timescale much shorter than that of thermodynamics, we first compute the elastic field within the body. We assume the mesa to be long enough as to neglect field variations along the $y$-direction, rendering our problem a two-dimensional (2D) one effectively.
We look for solutions of local equilibrium $\boldsymbol{\nabla}\cdot \vsig \pe \boldsymbol{0}$, in which the stress $\vsig$ is linearly and isotropically related to strain $\boldsymbol{\epsilon}$ via 
$\vsig \pe \frac{Y}{1+\nu}\left[\veps+\frac{\nu}{1-2\nu}\text{Tr}\left(\veps \right)\mathbb{1}\right]$ \cite{LandLifcEl}. For the sake of simplicity, we assume that the Young modulus $Y$ and the Poisson ratio $\nu$ are equal in film and mesa. In the 2D case, strain fields can be computed in an exact manner by utilizing the Airy function formulation (See, e.g.~\cite{Sadd}). Within this framework, the components of the stress are derived from a scalar function $\phi$ (the Airy function) such that  $\sigma_{xx} \pe \partial^2 \phi/\partial z^2$ (and $x \leftrightharpoons z$), $\sigma_{xz} \pe - \partial^2 \phi/\partial x \partial z$. Equilibrium is then satisfied when $\nabla^4\phi \equiv \frac{\partial^4\phi}{\partial x^4}+2\frac{\partial^4\phi}{\partial x^2\partial z^2}+\frac{\partial^4\phi}{\partial z^4} \pe 0$. In our mesa geometry we search for an exact solution for $\phi$ using a double Fourier transform in $x$ and $z$ \cite{Faux94}

\begin{strip}
\begin{align}
\label{airygood}
\rule{0mm}{15mm}
        \phi (x,z) = & \sum_{n=1}^{\infty}\cos (\beta_n x)\left[\left(A_n+C_n\frac{z}{H} \right)\frac{\sinh(\beta_n z)}{\beta_n^2 \, \sinh(\beta_n H)}
        +\left(D_n+E_n\frac{z}{H}\right)\frac{\cosh(\beta_n z)}{\beta_n^2 \, \cosh(\beta_n H)} \right] \nonumber \\
        & + \sum_{n=1}^{\infty} \sin(\alpha_n z)\left[
         F_n \frac{\cosh(\alpha_n x)}{\alpha_n^2\, \cosh (\alpha_n L)}
         + G_n \frac{x}{L} \frac{\sinh(\alpha_n x)}{\alpha_n^2 \, \sinh (\alpha_n L)} 
        \right] \nonumber \\
        & +\sum_{n=1}^{\infty}\cos(\alpha_n z)\left[H_n\frac{x}{L}\frac{\sinh(\alpha_n x)}{\alpha_n^2 \, \sinh(\alpha_n L)} + I_n \frac{\cosh(\alpha_n x)}{\alpha_n^2 \, \cosh(\alpha_n L)}\right] + J z^2 + K \frac{z^3}{H} 
        \, , 
\end{align}
\end{strip}
with $\alpha_n \pe n \pi / H$, $\beta_n \pe n \pi / L$. Here we have considered a coordinate system with origin in the center of the body, with $x$ ranging from $-L$ to $L$ (where $L \pe w/2$) and $z$ ranging from $-H$ to $H$ [with $H \pe (e+h)/2$]. The full solution to the elastic problem is subject to appropriate boundary conditions. The top and side facets are stress-free, i.~e., $\vsig\cdot \mathbf{n}_{z} \pe \mathbf{0}$ when $z \pe H$ and $\vsig\cdot \mathbf{n}_{x} \pe \mathbf{0}$ when $z \pe \pm L$ where $\mathbf{n}_\alpha$ is the unit vector in the direction $\alpha \pe x$ or $z$). We consider that the mesa is deposited on a rigid Si substrate that enforces zero displacement on the $z \pe -H$ lower surface \cite{JainHark95}. Growth conditions are such that the SiGe/Si interface is coherent, so that the force perpendicular to the interface ($\vsig \cdot \mathbf{n}_{z}$) and the displacement $\vu$ with respect to Si [accounting for the misfit $m \pe (a_{SiGe}-a_{Si})/a_{Si}$] are continuous across the boundary. To simplify our calculations, we use the decomposition $\vu \pe \bar{\vu} + \vu^*$ where $\bar{\vu}$ corresponds to an infinitely large ($w \rightarrow \infty$) flat film. In this special case, $\bar{\vu} \pe \vz$ in the mesa. In the film, however, we have $\bar{u} \pe 2 m \nu/(1-\nu) \left(0,0,z-(e-h)/2\right)$. Decomposing $\vu$ in this way produces a new unknown displacement vector $\vu^*$ that is effectively 2D, given its $y$-component is identically 0. This offers the possibility to make use of the aforementioned Airy formalism and tackle the problem in an analytical fashion. Furthermore, one can show that $\vu^*$ is the equilibrium displacement for a single body with its side facets being subjected to a step-like load with form 
$\sigma^*_{xx} \pe m Y/(1-\nu) \Theta\left[ z-(e-h)/2\right]$ and $\sigma^*_{xz} \pe 0$, where $x \pe \pm L$ and $\Theta$ is the Heaviside step function with center at the interface position. The other conditions are $\vsig^* \cdot \mathbf{n}_{z} \pe \vz$ when $z \pe H$, $\vu^* \pe  \vz$ when $z \pe -H$ and $\int_{-L}^L \epsilon_{zz}^* \, dx \pe 0$ when $z \pe -H$. The latter  expresses the fact that the total force upon the bottom surface ought to be zero. Imposing such boundary conditions upon the Airy function of Eq.~\eqref{airygood} gives rise to linear relations of the form 
$\sum_{m \pe 1}^\infty \left[ a_{n,m}^p A_m + c_{n,m}^p C_m + d_{n,m}^p D_m + \ldots + i_{n,m}^p I_m \right] + j_n^p J + k_n^p K \pe L_n$ with $p \pe 1, 2  \ldots 8$, $n \pe 1, 2, \ldots \infty$, plus 2 equations relating $J$ and $K$ to the other coefficients, where the  $a_{n,m}^p, c_{n,m}^p, \ldots i_{n,m}^p, j_n,  k_n, L_n$ are given as algebraic expressions involving $\tanh(\alpha_n L)$, \linebreak $\tanh(\beta_n H)$, $\sin(\alpha_n H)$, $e/H$ \cite{LongMesa}. 
We numerically solved these linear matrix equations by truncating series at a given order $N$ and inverting the linear relationships. This gives the values of the different coefficients $A_n, C_n, \ldots$ and thence the solution for the Airy function. With this solution in hand, we compute stresses and derive the strain tensor.


\subsection{Results}

\begin{figure*}[!ht] \centering
    \includegraphics[width=13cm]{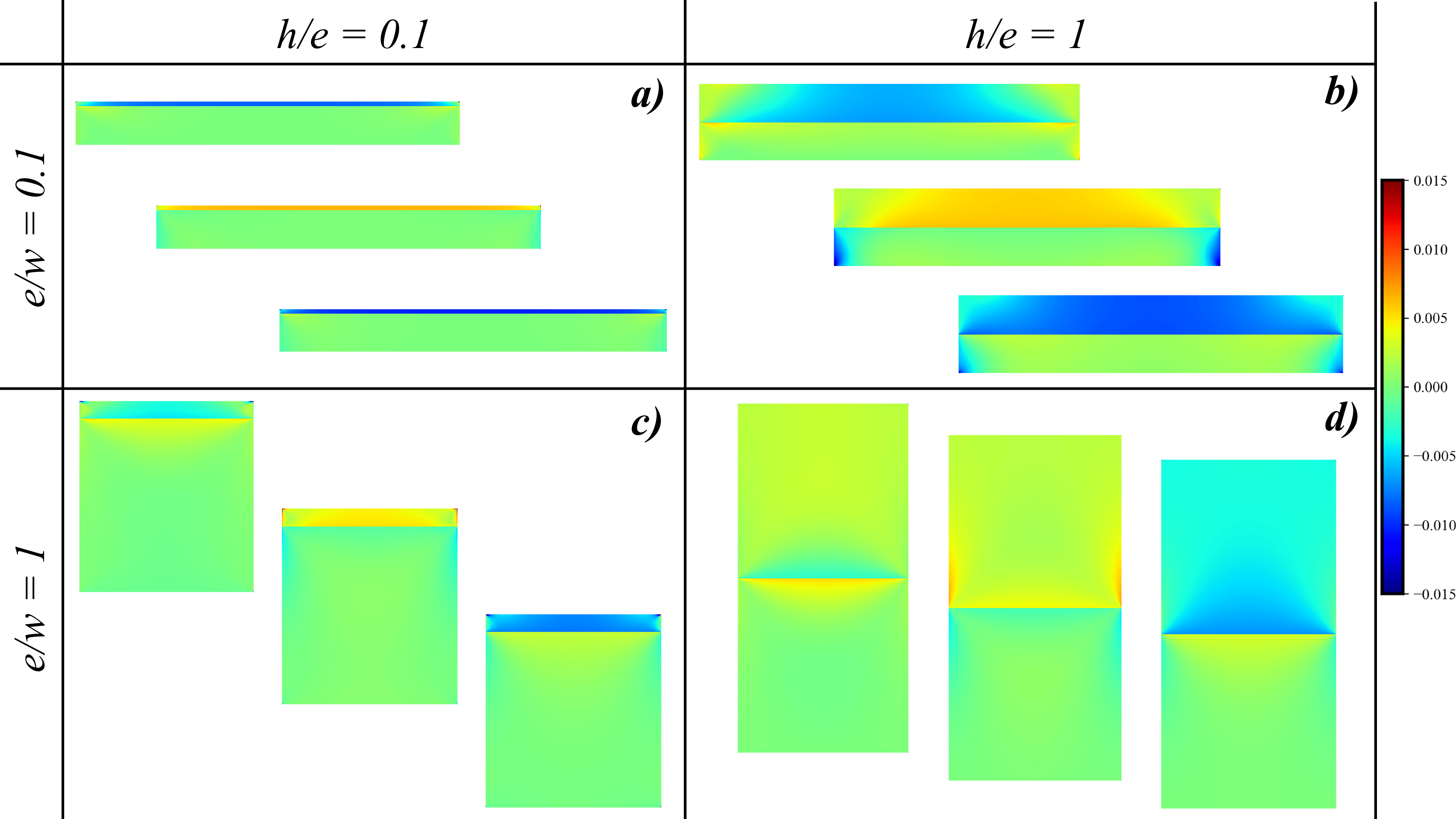}
    \caption{Strain maps for a different mesas with $e/w$ and $h/e \pe 1/10$ or $1$. For each geometry, we display $\epsilon_{xx}$ (up), $\epsilon_{zz}$ (middle) and $\rm{Tr}[\veps]$ (bottom) for a Si$_{0.8}$Ge$_{0.2}$ film on a Si mesa. The maps correspond to the solution of \eqref{airygood} with a truncation number $N \pe 15\,000$.}
    \label{fig-strain}
\end{figure*}

In Figure \ref{fig-strain} we show the resulting lateral strain map $\epsilon_{xx}$, the Poisson dilatation $\epsilon_{zz}$ and the total strain $\rm{Tr}[\veps]$ (where $\epsilon_{yy} \pe -m$). 
One can see that for small aspect ratios, $e/w \pe h/e \pe 1/10$, see Fig.~\ref{fig-strain}a, the system visibly resembles the thin film case, only deviating from this in the vicinity of the side facets. In this case, away from the facets, the film is nearly fully strained according to the substrate with the flat and infinite film displacement vector $\bar{\vu}$ corresponding to a biaxial compression in $x$ and $y$ ($\epsilon_{xx} \pe -m$) and a Poisson dilatation in $z$. The vicinity of the side facets is characterized by relaxation zones where the lateral strain evolves from some value depending on $z$ at the free facet to $-m$ in the middle of the mesa. This relaxation takes place over some length $\lambda$ that is a function of aspect ratios $e/w$ and $h/e$, see Fig.~\ref{fig-strain}. At the mesa edges, $\epsilon_{xx}$ vanishes and the film is relaxed in the $x$ direction, as can be experimentally measured in other similar geometries \cite{Bert18}. The relaxation zone near the edges characterizes the typical inhomogeneity of strain enforced by finite size effects and is at the core of the dynamical instability to be revealed in the following. For stronger finite size effects, with an aspect ratio $e/w \pe h/e \pe 1$, see Fig.~\ref{fig-strain}d, the two relaxation zones overlap considerably and the system, even in its midst, is far from resembling the full biaxial strain of the infinite film geometry. In this latter case,  strains $\epsilon_{xx}$ and $\epsilon_{zz}$ are relatively small within the film, leading to a reduced full strain $\rm{Tr}[\veps]$ (yet $\epsilon_{yy}$ is still $-m$). 
This kind of geometry is thus very suitable to effectively and efficiently relax the epitaxial stress and obtain highly relaxed deposits on substrates with detuned lattice parameters. 
These two cases (full strain with clearly separated relaxation zones \textit{vs} relaxed with overlapping relaxation zones, Figs.~\ref{fig-strain}a and d) are the two extremes when varying the geometric parameters ($e/w$ and $h/e$). 
For intermediate values of the different aspect ratios ($e/w$ and $h/e$), see Fig.~\ref{fig-strain}b and c, one finds that the relaxation zone increases when either $h/e$ increases for a given $e/w$ width ratio, or when $e/w$ decreases for a given $h/e$. It is noticeable that the full strain case is already achieved for values of $h/e$ and $e/w$ below typically 0.1, while the relaxed case occurs for values of the order or larger than 1. Hence, we can argue that relaxation has rather very localised effects on the edges of the mesa as soon as the aspect ratios $e/w$ and $h/e$ are both smaller than typically 0.1, whereas relaxation starts to significantly penetrate the mesa as soon as one of the aspect ratios is larger than typically 0.1. If, on the other hand, one seeks to relax the epitaxial stress almost completely and uniformly, one can see that this is already the case as soon as $h/e$ and $e/w$ are equal to or greater than 1, see Fig.~\ref{fig-strain}d. 
It means that a full strain cannot be maintained in a 50\,nm gate length. However, the relaxation remains very low when the raised Si is 10 times thicker than the epitaxial SiGe layer.

\begin{figure}[!ht] \centering
    \includegraphics[width=6.cm]{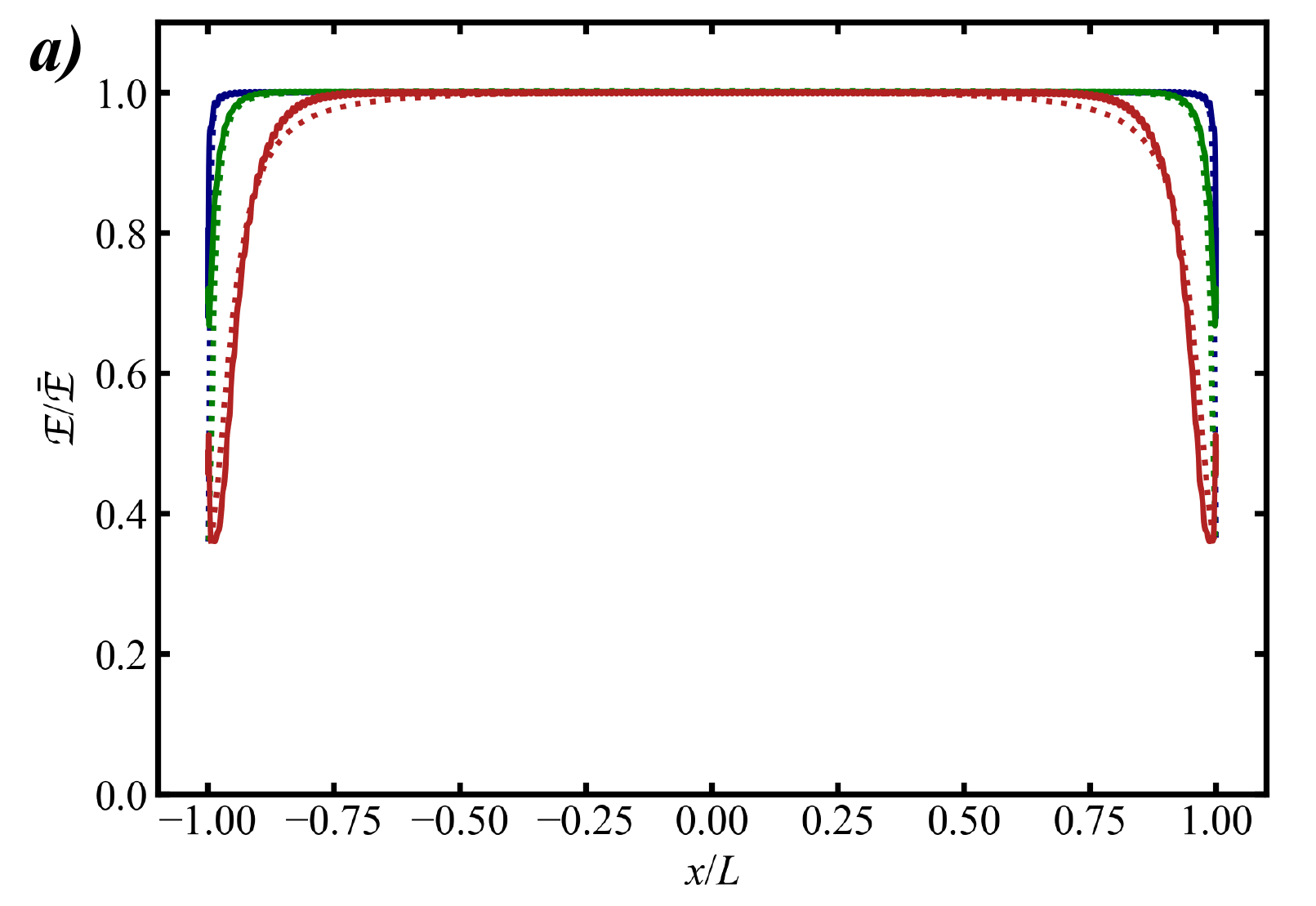}
    \includegraphics[width=6.cm]{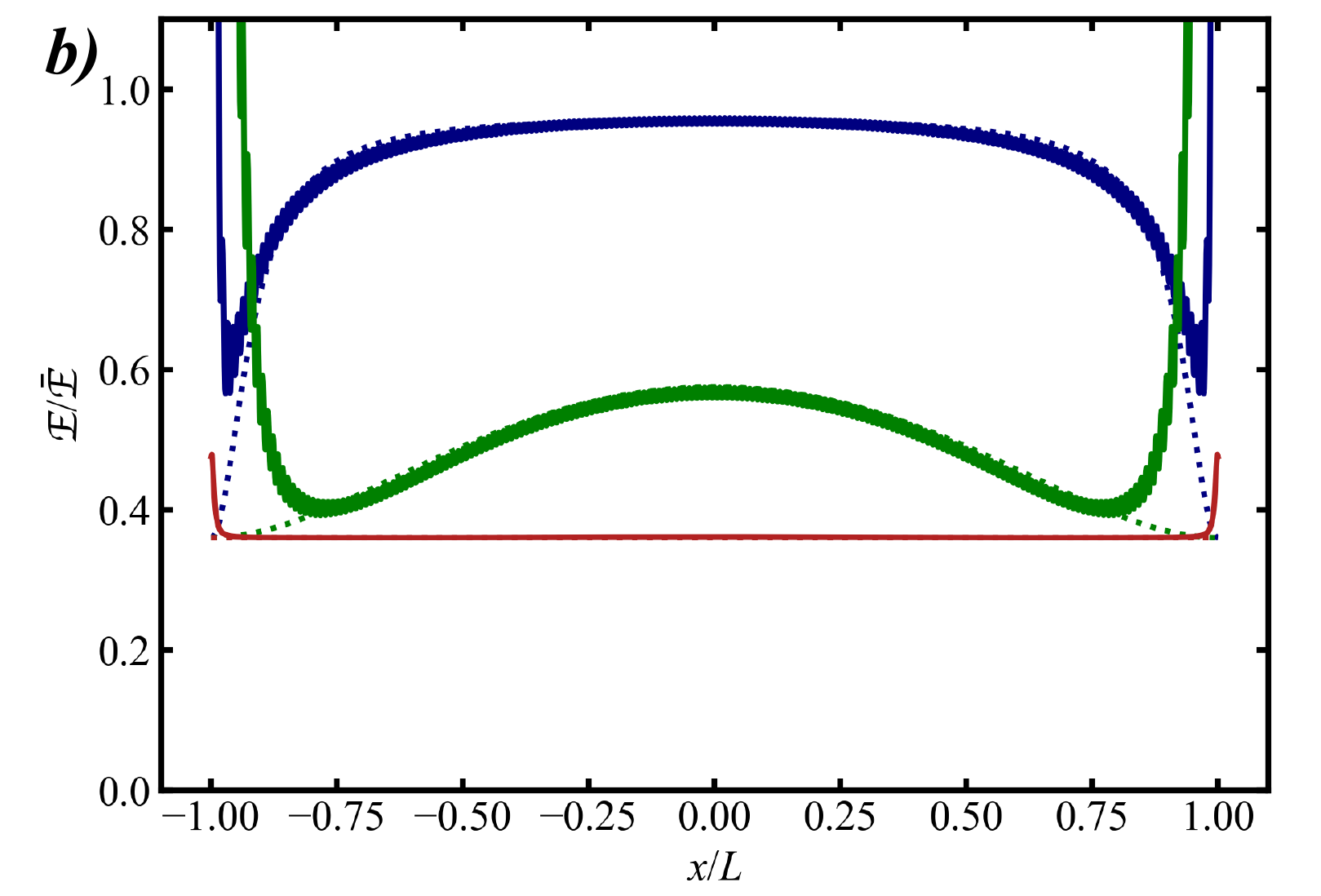}
    \caption{Energy density on top of the mesa $\mathcal{E}(x)/\bar{\mathcal{E}}$ (at $z\pe H$) as a function of $x/L$ in units of the maximum $\bar{\cE}$ for an infinite film for the full solution \eqref{airygood} of the elastic problem (solid line) and fit \eqref{fit} of this solution (dashed curve) for (a) $e/w=1/100$ and $h/e = 1/100$ (blue), $1/10$ (green) and $1$ (red)  and (b) $e/w = 1$ and same configurations. The fitting parameters are, for $e/w = 1/100$: $\le/L = 0.075$ and $A= 0.641$ for $h/e = 0.01$,  $\le/L = 0.15$ and $A= 0.643$  for $h/e = 0.1$, $\le/L = 0.31$ and $A= 0.647$ for $h/e = 1$ ; and for $e/w = 1$: $\le/L = 0.4$ and $A=0.61 $ for $h/e = 0.01$,  $\le/L = 1.5$ and $A= 1.3 $ for $h/e = 0.1$, $\le/L = 1.5$ and $A=0.061  $  for $h/e = 1$. (color online)
    }
    \label{fig-Eel}
\end{figure}

\begin{figure}[!ht] \centering
 \includegraphics[width=0.75\linewidth]{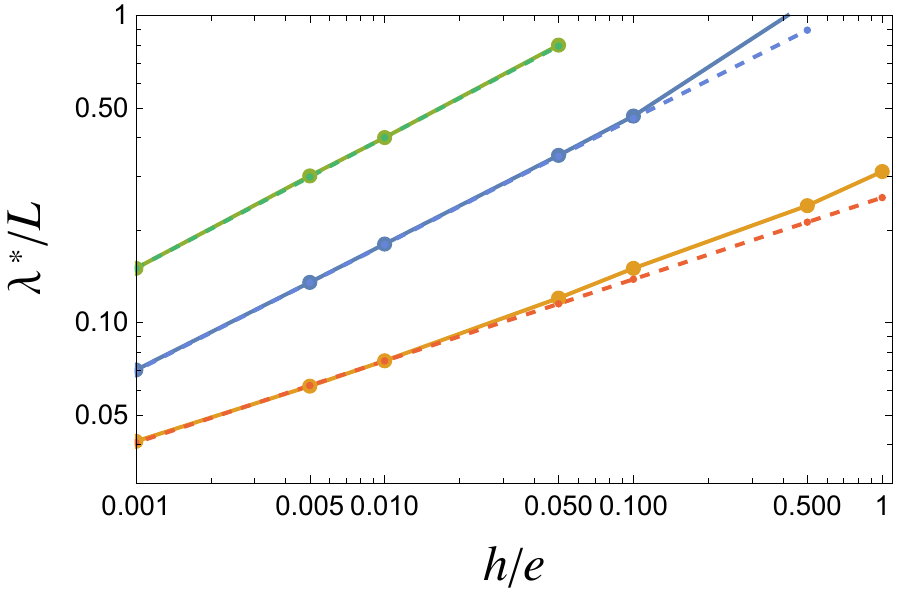}
  \caption{Typical size of the relaxation zone at the corners $\le/L$ as a function of the ratio between the film and the mesa thicknesses, $h/e$, for different mesa aspect ratio $e/w = 1$ (upper curve), $0.1$ (middle curve) and $0.01$ (lower curve). The dotted line is an algebraic fit $\le/L \varpropto (h/e)^\alpha$,  with $\alpha = 0.42$ (upper curve), $0.41$ (middle curve) and $0.27$ (lower curve). (We do not represent  values of $\le$ close to or greater than unity which are characterized by large uncertainties.) (color online)}
    \label{Fig4-fitparam}
\end{figure}

In order to characterize the inhomogeneity at the mesa surface, and for subsequent use in the dynamical analysis, we consider the elastic energy density $\cE \pe \fud \sigma_{i \, j} \epsilon_{i \, j}$ computed on the surface $z \pe H$. In Fig.~\ref{fig-Eel} we plot the typical evolution of this energy inhomogeneity for two aspect ratios $e/w$ but varying the film thickness $h$. 
The energy is maximum in the middle of the mesa where strain is higher, and it decreases towards the free border, where relaxation is maximum. For geometries that strain the film strongly, i.e., either thin enough films or sufficiently wide mesas (small aspect ratio $e/w$ and $h/e$), the energy density in the middle of the film and at the mesa surface is close to the value for a thin film, $\bcE \pe Y m^2/(1-\nu)$,  see Fig.~\ref{fig-Eel}a.
On the other hand, the film is clearly relaxed not only at its edges, but also in its middle in case the aspect ratios $e/w$ or $h/e$ are no longer small, see Fig.~\ref{fig-Eel}b.
Naturally, we can see from the elastic energy curves that relaxation is more efficient as the ratio $h/e$ increases for a given $e/w$ shape, see the different curves in Fig.~\ref{fig-Eel}a and b, but also it is more efficient as $e/w$ increases for a given $h/e$, see the comparison between Figs.~\ref{fig-Eel}a and b.   
At the corners ($x\pe \pm L, z \pe H$), the value of the energy density is dictated by the boundary conditions~: the elastic field must satisfy $\sigma_{xx} \pe \sigma_{zz} \pe \sigma_{xz} \pe 0$ so that $\epsilon_{xx} \pe \epsilon_{zz} \pe m \nu$ and $\sigma_{yy} \pe - m Y$ and eventually $\cE_c \pe \fud Y m^2 \pe \fud (1-\nu) \bcE$, which is $\simeq \! 0.36 \, \bcE$ for SiGe. This value is indeed well found by the numerical calculation at or near the corners, except in some conditions where the usual Gibbs effect can be detected 
\footnote{We can observe a numerical blow-up at the exact corners in some conditions, see Figs.~\ref{fig-strain} and \ref{fig-Eel}. This has been observed previously by fellow workers using different computational methods (see \cite{JainHark95} using finite element methods), and is nothing but the Gibbs effect when the Fourier decomposition converges only very slowly to the exact solution when a discontinuity is present. It is here an undesirable and non-physical effect that limits the application of the Fourier calculation. It is hardly visible in Fig.~\ref{fig-Eel}a) for small $e/w$, but is clearly at work at the corners in  Fig.~\ref{fig-Eel}b) when $e/w \pe 1 $ and when $h/e$ is not too small.}.
Utilizing the boundary conditions similarly, we are able to deduce that the energy density profile at the corners satisfies $d \cE / dx (x\pe \pm L, z\pe H) = 0$ which is also verified by the full solution (except for the numerical singularity at the corners). For later use in the dynamical analysis, we fit this energy profile with the following form
\begin{multline}
\label{fit}
 \cE(x) = \bcE \left[ \fud (1-\nu)  + f(x) \right] \mbox{ with } \\
 f(x) =  A( \mbox{$\frac{e}{w}, \frac{h}{e}$} ) \frac{\left( \dfrac{L^2-x^2}{\le^2}\right)^2}{1+\left( \dfrac{L^2-x^2}{\le^2}\right)^2 }  \, , 
\end{multline}
with two fitting parameters, the amplitude $A( \mbox{$\frac{e}{w}, \frac{h}{e}$})$ and the relaxation zone width $\le$. This fit is plotted as a dotted line in Fig.~\ref{fig-Eel}, and is relatively close to the Fourier series calculation except at the corners when the Gibbs effect is non negligible, where on the contrary, we impose on the fit to tend towards the exact analytical value $\cE_c / \bcE \pe \fud (1-\nu)$. We plot in Fig.~\ref{Fig4-fitparam} the typical width of the relaxation zone $\le$ as a function of the system geometric parameters, $h/e$ and $e/w$. We find that the relaxation zone width $\le$ naturally increases when the film height increases, with a typical algebraic behavior. It is noticeable that its exponent is similar for $e/w \pe 0.1$ and 1, but departs from the fully strain case when $e/w \pe 0.01$. Finally, the relaxation zone also increases with the mesa aspect ratio $e/w$ as finite size effects in the mesa increase. 


\section{Dynamical evolution}
\subsection{Surface diffusion framework and eigenmodes}

Thanks to the above solution for the elastic field, we now turn to the dynamical evolution of the mesa instability in the presence of such an energy inhomogeneity. We consider the usual thermodynamic framework in which surface diffusion occurs only on the mesa upper surface described by $z \pe h(x,t)$ \cite{SpenVoor91}. We therefore neglect mass transport across the mesa corners. Mass conservation thence requires $\partial h / \partial \, t \pe - \nabla \cdot \mathbf{J}$ where $\mathbf{J}$ is the surface current density. This relates to the surface chemical potential via $\mathbf{J} \pe - D \nabla \mu$, with some diffusion coefficient $D$ \footnote{We assume some slopes $\frac{\partial h}{\partial X} \ll 1$ so that surface gradients may be approximated by gradients}. The chemical potential at the surface relates to the capillary and elastic energies as $\mu \pe - \gamma \Delta h + \cE(x,t)$ in the small-slope approximation. In the following, we assume that the energy profile $\cE(x,t)$ is set by the finite-size effects that we calculated above, and that it does not change throughout the dynamical evolution, i.e.~$\cE(x,t)=\cE(x)$. This assumption is supported by the fact that (i) the extra relaxation enforced by the morphological evolution of the free surface is rather small (see \cite{AquaBerb13} for the infinite substrate case) and that (ii) we will consider the initial evolution of the evolution. During these early stages the slope and the concomitant surface relaxation are relatively small. In contrast, the driving force associated to finite-size effects is noticeable and is present from the very beginning of the evolution. The morphological instability thence occurs in a given energetic profile. It is a noticeable difference with the strain-induced ATG instability \cite{AsarTill72,Grin86,AquaBerb13} or quantum dot nucleation \cite{GailAqua13,LiuBerb17} where on the contrary, the energetic profile is precisely given by the corrugation of the surface and proportional to $h(x,t)$ \footnote{We typically consider the inhomogeneity given by \eqref{fit} but the following results are quite insensitive to this exact form as long as the energy profile is similar.}. In this framework, 
we use the scaling $x \pe \le \, \xt$, $h \pe \ls \hti$, $t \pe \ts \tt$ with 
\begin{equation}
\label{scales} 
\ls = \le^2 \bcE / \gamma \quad \mbox{and} \quad \ts = \le^4 / D \gamma \, , 
\end{equation} 
so that the evolution equation takes the form 
\begin{equation}
\label{evol} 
\frac{\partial \hti}{\partial \tt} + \frac{\partial^4 \hti}{\partial \xt^4} = \frac{\partial^2 f}{\partial \xt^2}  \, , 
\end{equation} 
with the elastic inhomogeneity $f(x)$ given in Eq.~\eqref{fit}. It is noteworthy that the scales of the mesa instability \eqref{scales} are again different to those of the ATG instability that are $\lambda_{\ATG} \pe \gamma / \bcE$ and $\tau_{\ATG} \pe \lambda_{\ATG}^4/D \gamma$. Contrarily to $\lambda_{\ATG} \varpropto 1/m^2$, here $\le$ does not depend on $m$ and is just a function of the geometric parameters $h$, $e$ and $w$. As an example, and in order to give some orders of magnitude relevant to the experiments, we display in Table~\ref{table} the values of the time and space scales for a Si$_{0.8}$Ge$_{0.2}$ film on Si at $T\pe 550^\circ$C. We first notice that these scales explicitly depend on the geometry of the system. We also note that the wider system, the bigger the time scale $\tau^*$ is, so that naturally, the evolution occurs much faster on a small mesa compared to a big one, all aspect ratios being equal. The scales $\lambda^*$ and $l^*$ also decrease proportionally to $w$, so that the amplitude of the beading at the mesa edges is expected to decrease on a narrow mesa.


\begin{table*}[ht!] \centering
\begin{tabularx}{12.3cm}{|c||c|c||c|c||c|} 
   \hline
     & \multicolumn{2}{c|}{$e/w= 1/100 \, ; \, h/e= 1$ \,} \rule{0mm}{3mm} & \multicolumn{2}{|c||}{$e/w= 1 \, ; \, h/e= 1/100$} &  \\
    \cline{2-6} 
    & $w=200$\,nm &  $w=20$\,nm & $w=200$\,nm & $w=20$\,nm &  \rule{0mm}{3mm} \\
   \cline{1-5} 
  $\lambda^*$\, (nm) & \rule{0mm}{3mm}  31 & 3.1 & 40 & 4 &  $\lambda_{ATG} \pe 43$\,nm  \\
     \cline{1-5} 
  $l^*$\, (nm)  & \rule{0mm}{3mm}  8.7 & 0.09 & 14 & 0.01 & \\
     \cline{1-5} 
  $\tau^*$\, (s) & \rule{0mm}{3mm}  210 & 0.021 & 580 & 0.058 &  $\tau_{\ATG} \pe 780$\,s \\
    \hline
\end{tabularx}
\caption{Examples of the characteristic time and space scales for the elastic relaxation and ATG instability corresponding to a Si$_{0.8}$Ge$_{0.2}$ film on Si at $T\pe 550^\circ$C \cite{AquaGouy13}.} 
\label{table} 
\end{table*}

It is possible to solve the evolution equation \eqref{evol} with the initial condition $h(x,0) = H$ utilizing standard methods of functional analysis in presence of the forcing term related to the energy inhomogeneity (see a similar problem in \cite{ZhenHann19}). We assume the continuity of $\mu$ (neglecting lateral corrugation) and neglect mass flow at the corners, so that we enforce $\partial^2 \hti / \partial \xt^2 \pe 0$ and $\partial^3 \hti / \partial \xt^3 \pe 0$ when $\xt \pe \pm \Lt$ where $\Lt \pe L / \le$. One may first compute the stationary solution $\hs(\xt)$ of Eq.~\eqref{evol} given by 
\begin{equation} 
\label{hs}
\hs (\xt) = I_2 (\xt)  + \Ht - \frac{1}{\Lt} \int_0^{\Lt} d\xt \, I_2(\xt) \, , 
\end{equation}
where $I_2(\xt) \pe \int_0^{\xt} du \, \int_0^u dv \, f(v,\Lt)$. Having this long-time stationary solution, one can find $\hti$ thanks to the decomposition $\hti \pe \hs(\xt) - \hu (\xt,\tt)$ where $\hu$ satisfies the Schr\"odinger equation 
\begin{equation} 
\label{Schro}
\frac{\partial \hu}{\partial \, \tt} = - \cH [\hu] \quad \mbox{with} \quad \cH \equiv \frac{\partial^4}{\partial \xt^4} \, . 
\end{equation} 
The eigenfunctions $\phi_n$ of $\cH$ such as $\cH[\phi_n] \pe E_n \phi_n$, $n\pe 1, 2 \ldots$, are merely 
\begin{equation} 
\label{eigenfn}
\phi_n(\xt) = \frac{1}{\sqrt{\Lt}} \frac{\cosh(k_n \Lt) \cos(k_n \xt) + \cos(k_n \Lt) \cosh(k_n \xt)}{\sqrt{\cos^2(k_n \Lt) + \cosh^2(k_n \Lt)}} \, , 
\end{equation} 
where $k_n \pe E_n^{1/4}$ satisfies the characteristic equation originating from the boundary conditions 
\begin{equation} 
\label{carac}
\tan (k_n \Lt) + \tanh(k_n \Lt) = 0 \, ,
\end{equation}
whose roots can only be found numerically. Yet, a very good approximation of them is $k_n \Lt \! \simeq \!  (n+3/4) \pi$. Decomposing $h_1$ in terms of these eigenfunctions, one eventually finds the solution to the surface dynamics as
\begin{equation}
	\label{solution}
\hti (\xt,\tt) = \hs(\xt) - \sum_{n\pe 1}^\infty c_n e^{-k_n^4 \tt} \, \phi_n(\xt) \, , 
\end{equation}
where the coefficients $c_n$ are given by the initial condition 
$c_n \pe \int_{-\Lt}^{\Lt} d\xt\, \phi_n(\xt) \hs(\xt)$.


\subsection{Results}

\begin{figure}[!ht] \centering
    \includegraphics[width=6cm]{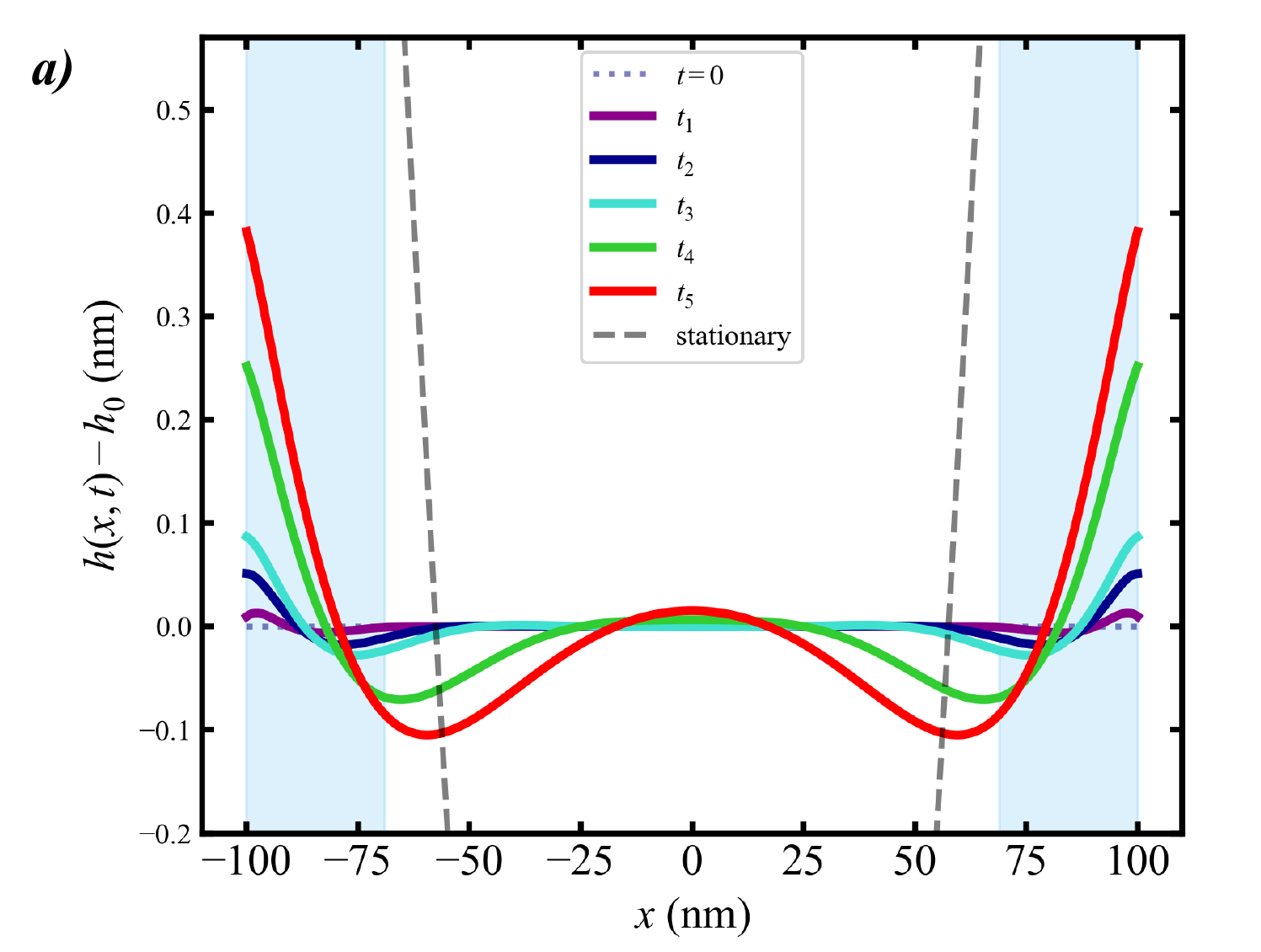}
    \includegraphics[width=6cm]{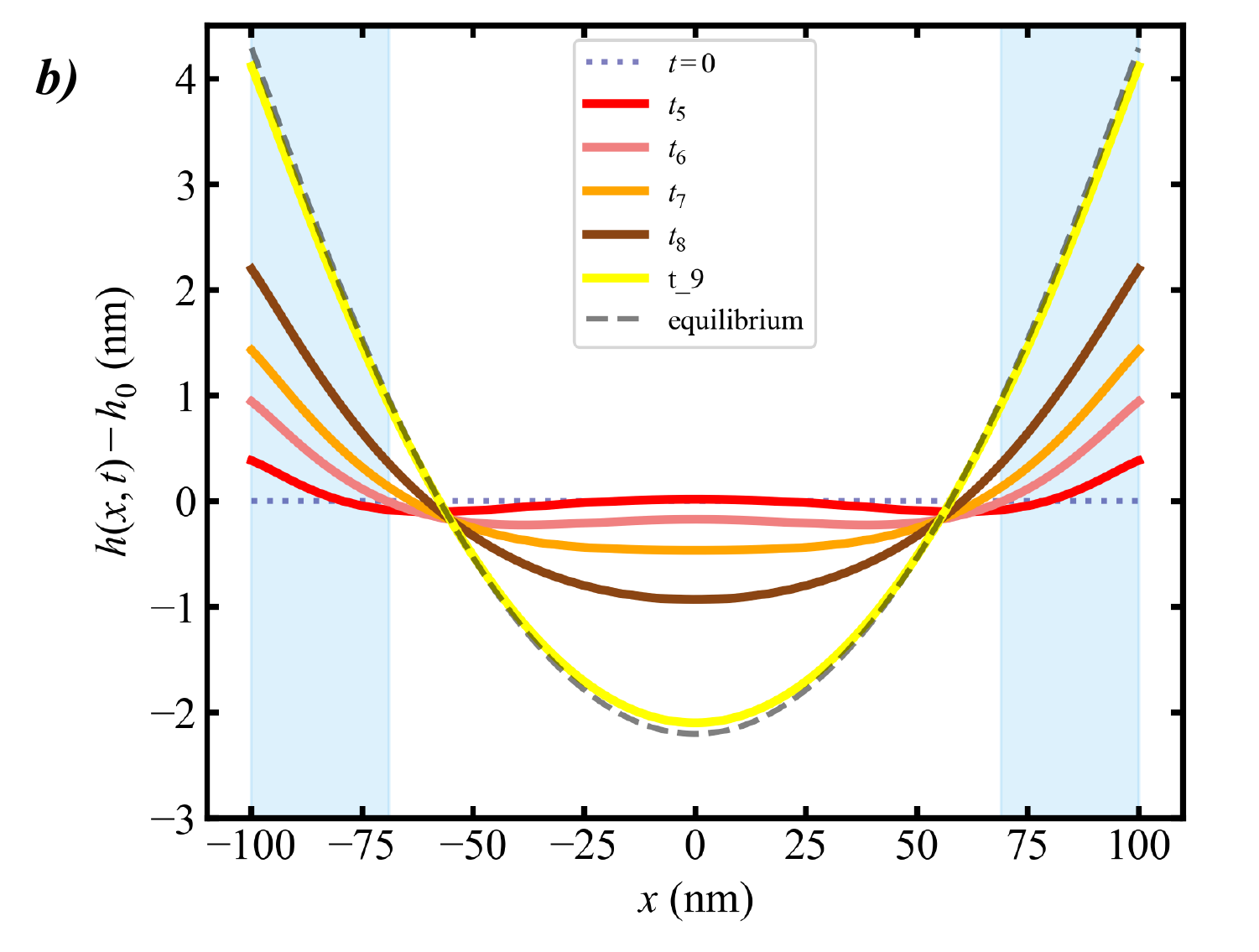}
    \caption{Evolution of the mesa surface driven by the finite-size elastic relaxation for a Si$_{0.8}$Ge$_{0.2}$ film on a Si mesa with $e\pe h \pe 2$\,nm and $w \pe 200$\,nm for the initial (a) and final (b) times, for $t_1 \pe 10^{-5}$, $t_2 \pe 5\, 10^{-5}$, $t_3 \pe 10^{-4}$, $t_4 \pe 5\, 10^{-4}$, $t_5 \pe 10^{-3}$, $t_6 \pe  5\, 10^{-3} $, $t_7 \pe  10^{-2}$, $t_8 \pe 2\,  10^{-2}$, $ t_9 \pe 10^{-1}$. The dynamics follows the solution \eqref{solution} and the elastic energy profile is characterized by $\le \pe 31$\,nm and $A \pe 0.647$ for this geometry. The equilibrium stationary solution (dashed line) corresponds to Eq.~\eqref{hs}. The typical relaxation zones (where strain is significantly different from the infinite-mesa limit) are shaded in blue. (color online)
}
    \label{Fig5-evol-largew}
\end{figure}

The resulting evolution described by \eqref{solution} is plotted in Fig.~\ref{Fig5-evol-largew}. 
%
Firstly, we note that the elastic density gradient  drives mass transfer toward the sides of the mesa, where the elastic energy is lower. At the beginning of the relaxation, we observe that mass transport initiates within the relaxation zones, where there is a noticeable gradient in elastic energy. Meanwhile, the middle of the mesa remains mostly stationary and flat. The mesa is then W-shaped with a flat central area that gradually narrows. In latter stages, however, the evolution progressively spreads in toward the middle of the mesa. This is the response of the surface capillarity, which tends to minimize the measure of mesa surface area facing the vacuum. In the long run, the surface corrugates throughout the top surface and is characterized by mass transfer from regions close to the middle of the mesa towards its ends. At this point, matter is effectively transferred from the middle of the mesa toward its sides, quickly giving rise to the equilibrium V-shape. 
 
It is noticeable that the evolution on the different geometries occurs on different time scales. If Fig.~\ref{Fig5-evol-largew} is characterized by dimensionless time scales, for large mesas, the instability occurs over a rather long time scale, while it occurs over short time scale for compact mesas. For example, for the aspect ratios $e/w \pe 1/100$ and $h/e \pe 1$ of Fig.~\ref{Fig5-evol-largew}, one finds for a Si$_{0.8}$Ge$_{0.2}$ film on a Si mesa with $w \pe 200$\,nm at $T\pe 550^\circ$C, $\ts \pe 210$\,s while it is just $0.021$s when $w\pe 20$\,nm. Consequently, in the latter case, the final stationary V shape is quickly reached, while it is clearly not the case for the larger mesas. Note also that the counter-acting driving force of the capillarity is lessened on larger mesas due to lower surface-to-volume ratio on these. The theoretical framework given here relies on different assumptions (static elastic field, small-slope approximation, absence of mass transfer from the sides ...), but we argue that the main physical picture is captured by this description~: mass transfer to the mesa ends where the elastic energy is lower, with a (transitory) W shape or (stationary) V shape and speed that are function of the mesa aspect ratios. This study therefore allows us to exhibit this new mode of relaxation on a minimal model including the crucial effects, and paves the way for further studies incorporating other effects that may come into play, such as crystal anisotropy \cite{AquaGouy11,AquaGouy13} or the competition between finite size relaxation and morphology-induced relaxation.

\section{Conclusions} 

We have investigated the elastic strain relaxation and the associated morphological evolution for SiGe stressors in small gate MOSFET devices. We have characterized a new morphological relaxation that drives mass transfer towards the edges of a finite strained nanolayers deposited on a mesa. Its dynamics originates from strain inhomogeneity near the free faces of the system that occurs over a relaxation zone that is function of the film thickness and mesa aspect ratio. The evolution displays different shapes and speed as a function of the mesa geometric parameters and may eventually lead to a stationary profile when capillarity counterbalances elasticity. This analysis reveals general mechanisms expected to be at work in several device geometries and also well beyond this framework. It opens the way for a better understanding and control of strain-engineering using finite-size effects experienced when using SiGe tressors. It should help to optimize the design of new geometries intended to improve the materials properties (mobility). It is currently investigated for the development of innovative, efficient and reliable systems for the pursuit of miniaturization in microelectronics, typically for node 28nm BiCMOS transistors and beyond.

\bibliographystyle{elsarticle-num} 

\begin{thebibliography}{10}
\expandafter\ifx\csname url\endcsname\relax
  \def\url#1{\texttt{#1}}\fi
\expandafter\ifx\csname urlprefix\endcsname\relax\def\urlprefix{URL }\fi
\expandafter\ifx\csname href\endcsname\relax
  \def\href#1#2{#2} \def\path#1{#1}\fi

\bibitem{FeraColi11}
I.~Ferain, C.~A. Colinge, J.-P. Colinge,
  \href{https://doi.org/10.1038/nature10676}{Multigate transistors as the
  future of classical metal--oxide--semiconductor field-effect transistors},
  Nature \textbf{479}~(7373) (2011) 310--316.
\newblock \href {http://dx.doi.org/10.1038/nature10676}
  {\path{doi:10.1038/nature10676}}.
\newline\urlprefix\url{https://doi.org/10.1038/nature10676}

\bibitem{Brue95}
M.~Bruel, Silicon on insulator material technology, Electronics Letters
  \textbf{31} (1995) 1201--1202(1).

\bibitem{WebeJosse14}
O.~Weber, E.~Josse, F.~Andrieu, A.~Cros, E.~Richard, P.~Perreau, E.~Baylac,
  N.~Degors, C.~Gallon, E.~Perrin, S.~Chhun, E.~Petitprez, S.~Delmedico,
  J.~Simon, G.~Druais, S.~Lasserre, J.~Mazurier, N.~Guillot, E.~Bernard,
  R.~Bianchini, L.~Parmigiani, X.~Gerard, C.~Pribat, O.~Gourhant, F.~Abbate,
  C.~Gaumer, V.~Beugin, P.~Gouraud, P.~Maury, S.~Lagrasta, D.~Barge, N.~Loubet,
  R.~Beneyton, D.~Benoit, S.~Zoll, J.-D. Chapon, L.~Babaud, M.~Bidaud,
  M.~Gregoire, C.~Monget, B.~Le-Gratiet, P.~Brun, M.~Mellier, A.~Pofelski,
  L.~Clement, R.~Bingert, S.~Puget, J.-F. Kruck, D.~Hoguet, P.~Scheer,
  T.~Poiroux, J.-P. Manceau, M.~Rafik, D.~Rideau, M.-A. Jaud, J.~Lacord,
  F.~Monsieur, L.~Grenouillet, M.~Vinet, Q.~Liu, B.~Doris, M.~Celik,
  S.~Fetterolf, O.~Faynot, M.~Haond, 14nm fdsoi technology for high speed and
  energy efficient applications, in: 2014 Symposium on VLSI Technology
  (VLSI-Technology): Digest of Technical Papers, 2014, pp. 1--2.
\newblock \href {http://dx.doi.org/10.1109/VLSIT.2014.6894343}
  {\path{doi:10.1109/VLSIT.2014.6894343}}.

\bibitem{BaeBae16}
D.-i. Bae, G.~Bae, K.~K. Bhuwalka, S.-H. Lee, M.-G. Song, T.-s. Jeon, C.~Kim,
  W.~Kim, J.~Park, S.~Kim, U.~Kwon, J.~Jeon, K.-J. Nam, S.~Lee, S.~Lian, K.-i.
  Seo, S.-G. Lee, J.~H. Park, Y.-C. Heo, M.~S. Rodder, J.~A. Kittl, Y.~Kim,
  K.~Hwang, D.-W. Kim, M.-s. Liang, E.~Jung, A novel tensile si (n) and
  compressive sige (p) dual-channel cmos finfet co-integration scheme for 5nm
  logic applications and beyond, in: 2016 IEEE International Electron Devices
  Meeting (IEDM), 2016, pp. 28.1.1--28.1.4.
\newblock \href {http://dx.doi.org/10.1109/IEDM.2016.7838496}
  {\path{doi:10.1109/IEDM.2016.7838496}}.

\bibitem{RiegVogl93}
M.~M. Rieger, P.~Vogl,
  \href{https://link.aps.org/doi/10.1103/PhysRevB.48.14276}{Electronic-band
  parameters in strained Si$_{1-x}$Ge$_{x}$ alloys on Si$_{1-y}$Ge$_{y}$
  substrates}, Phys. Rev. B \textbf{48} (1993) 14276--14287.
\newblock \href {http://dx.doi.org/10.1103/PhysRevB.48.14276}
  {\path{doi:10.1103/PhysRevB.48.14276}}.
\newline\urlprefix\url{https://link.aps.org/doi/10.1103/PhysRevB.48.14276}

\bibitem{NayaChun94}
D.~K. Nayak, S.~K. Chun, \href{https://doi.org/10.1063/1.111558}{Low‐field
  hole mobility of strained si on (100) Si$_{1-x}$Ge$_{x}$ substrate}, Appl.
  Phys. Lett. \textbf{64}~(19) (1994) 2514--2516.
\newblock \href {http://arxiv.org/abs/https://doi.org/10.1063/1.111558}
  {\path{arXiv:https://doi.org/10.1063/1.111558}}, \href
  {http://dx.doi.org/10.1063/1.111558} {\path{doi:10.1063/1.111558}}.
\newline\urlprefix\url{https://doi.org/10.1063/1.111558}

\bibitem{GhanArms03}
T.~Ghani, M.~Armstrong, C.~Auth, M.~Bost, P.~Charvat, G.~Glass, T.~Hoffmann,
  K.~Johnson, C.~Kenyon, J.~Klaus, B.~McIntyre, K.~Mistry, A.~Murthy,
  J.~Sandford, M.~Silberstein, S.~Sivakumar, P.~Smith, K.~Zawadzki,
  S.~Thompson, M.~Bohr, A 90nm high volume manufacturing logic technology
  featuring novel 45nm gate length strained silicon cmos transistors, in: IEEE
  International Electron Devices Meeting 2003, 2003, pp. 11.6.1--11.6.3.
\newblock \href {http://dx.doi.org/10.1109/IEDM.2003.1269442}
  {\path{doi:10.1109/IEDM.2003.1269442}}.

\bibitem{ThomArms04}
S.~Thompson, M.~Armstrong, C.~Auth, S.~Cea, R.~Chau, G.~Glass, T.~Hoffman,
  J.~Klaus, Z.~Ma, B.~Mcintyre, A.~Murthy, B.~Obradovic, L.~Shifren,
  S.~Sivakumar, S.~Tyagi, T.~Ghani, K.~Mistry, M.~Bohr, Y.~El-Mansy, A logic
  nanotechnology featuring strained-silicon, IEEE Electron Device Letters
  {\textbf{25}}~(4) (2004) 191--193.
\newblock \href {http://dx.doi.org/10.1109/LED.2004.825195}
  {\path{doi:10.1109/LED.2004.825195}}.

\bibitem{BerbAqua14}
I.~Berbezier, J.-N. Aqua, M.~Aouassa, L.~Favre, S.~Escoubas, A.~Gouy{\'e},
  A.~Ronda,
  \href{http://link.aps.org/doi/10.1103/PhysRevB.90.035315}{Accommodation of
  sige strain on a universally compliant porous silicon substrate}, Phys. Rev.
  B {\bf{90}} (2014) 035315.
\newline\urlprefix\url{http://link.aps.org/doi/10.1103/PhysRevB.90.035315}

\bibitem{AquaFavr15}
J.-N. Aqua, L.~Favre, A.~Ronda, A.~Benkouider, I.~Berbezier,
  \href{http://dx.doi.org/10.1021/acs.cgd.5b00485}{Configurable compliant
  substrates for sige nanomembrane fabrication}, Cryst. Growth Des.
  {\textbf{15}} (2015) 3399.
\newblock \href {http://dx.doi.org/10.1021/acs.cgd.5b00485}
  {\path{doi:10.1021/acs.cgd.5b00485}}.
\newline\urlprefix\url{http://dx.doi.org/10.1021/acs.cgd.5b00485}

\bibitem{Hu91}
S.~M. Hu, \href{https://doi.org/10.1063/1.349282}{Stress‐related problems in
  silicon technology}, J. Appl. Phys. \textbf{70}~(6) (1991) R53--R80.
\newblock \href {http://arxiv.org/abs/https://doi.org/10.1063/1.349282}
  {\path{arXiv:https://doi.org/10.1063/1.349282}}, \href
  {http://dx.doi.org/10.1063/1.349282} {\path{doi:10.1063/1.349282}}.
\newline\urlprefix\url{https://doi.org/10.1063/1.349282}

\bibitem{Faux94}
D.~A. Faux, \href{https://doi.org/10.1063/1.355881}{The fourier series method
  for the calculation of strain relaxation in strained‐layer structures}, J.
  Appl. Phys. \textbf{75}~(1) (1994) 186--192.
\newblock \href {http://arxiv.org/abs/https://doi.org/10.1063/1.355881}
  {\path{arXiv:https://doi.org/10.1063/1.355881}}, \href
  {http://dx.doi.org/10.1063/1.355881} {\path{doi:10.1063/1.355881}}.
\newline\urlprefix\url{https://doi.org/10.1063/1.355881}

\bibitem{JainHark95}
S.~C. Jain, A.~H. Harker, A.~Atkinson, K.~Pinardi,
  \href{https://doi.org/10.1063/1.360257}{Edge‐induced stress and strain in
  stripe films and substrates: A two‐dimensional finite element calculation},
  J. Appl. Phys. \textbf{78}~(3) (1995) 1630--1637.
\newblock \href {http://arxiv.org/abs/https://doi.org/10.1063/1.360257}
  {\path{arXiv:https://doi.org/10.1063/1.360257}}, \href
  {http://dx.doi.org/10.1063/1.360257} {\path{doi:10.1063/1.360257}}.
\newline\urlprefix\url{https://doi.org/10.1063/1.360257}

\bibitem{LandLifcEl}
L.~Landau, E.~Lifchitz, {Theory of Elasticity}, USSR Academy of Sciences,
  Moscow, USSR, 1986.

\bibitem{Sadd}
M.~H. Sadd, Elasticity, Theory, Applications and Numerics, Elsevier, 2005.

\bibitem{LongMesa}
K.~Rodriguez~Hannikainen, J.-N. Aqua, to be published.

\bibitem{Bert18}
R.~Berthelon, Strain integration and performance optimization in sub-20nm fdsoi
  cmos technology, Ph.D. thesis, Univ. Toulouse 3 Paul Sabatier (2018).

\bibitem{SpenVoor91}
B.~J. Spencer, P.~W. Voorhees, S.~H. Davis,
  \href{http://link.aps.org/doi/10.1103/PhysRevLett.67.3696}{Morphological
  instability in epitaxially strained dislocation-free solid films}, Phys. Rev.
  Lett. 67 (1991) 3696--3699.
\newblock \href {http://dx.doi.org/10.1103/PhysRevLett.67.3696}
  {\path{doi:10.1103/PhysRevLett.67.3696}}.
\newline\urlprefix\url{http://link.aps.org/doi/10.1103/PhysRevLett.67.3696}

\bibitem{AquaBerb13}
J.-N. Aqua, I.~Berbezier, L.~Favre, T.~Frisch, A.~Ronda,
  \href{http://www.sciencedirect.com/science/article/pii/S0370157312002761}{Growth
  and self-organization of {SiGe} nanostructures}, Phys. Rep. 522 (2013) 59 --
  189.
\newblock \href
  {http://dx.doi.org/http://dx.doi.org/10.1016/j.physrep.2012.09.006}
  {\path{doi:http://dx.doi.org/10.1016/j.physrep.2012.09.006}}.
\newline\urlprefix\url{http://www.sciencedirect.com/science/article/pii/S0370157312002761}

\bibitem{AsarTill72}
R.~J. Asaro, W.~A. Tiller, Interface morphology developpment during
  stress-corrosion cracking: Part 1. {V}ia surface diffusion, Metall. Trans.
  {\bf 3} (1972) 1789.

\bibitem{Grin86}
M.~A. Grinfeld, Instability of the separation boundary between a
  nonhydrostatiscally stressed elastic body and a melt, Sov. Phys. Dokl. {\bf
  31} (1986) 831.

\bibitem{GailAqua13}
P.~Gaillard, J.-N. Aqua, T.~Frisch,
  \href{http://link.aps.org/doi/10.1103/PhysRevB.87.125310}{Kinetic monte carlo
  simulations of the growth of silicon germanium pyramids}, Phys. Rev. B 87
  (2013) 125310.
\newline\urlprefix\url{http://link.aps.org/doi/10.1103/PhysRevB.87.125310}

\bibitem{LiuBerb17}
K.~Liu, I.~Berbezier, T.~David, L.~Favre, A.~Ronda, M.~Abbarchi, P.~W.
  Voorhees, J.-N. Aqua,
  \href{https://link.aps.org/doi/10.1103/PhysRevMaterials.1.053402}{Nucleation
  versus instability race in strained films}, Phys. Rev. Materials 1 (2017)
  053402.
\newblock \href {http://dx.doi.org/10.1103/PhysRevMaterials.1.053402}
  {\path{doi:10.1103/PhysRevMaterials.1.053402}}.
\newline\urlprefix\url{https://link.aps.org/doi/10.1103/PhysRevMaterials.1.053402}

\bibitem{AquaGouy13}
J.-N. Aqua, A.~Gouy{\'e}, A.~Ronda, T.~Frisch, I.~Berbezier,
  \href{http://link.aps.org/doi/10.1103/PhysRevLett.110.096101}{Interrupted
  self-organization of {SiGe} pyramids}, Phys. Rev. Lett. {\bf 110} (2013)
  096101.
\newline\urlprefix\url{http://link.aps.org/doi/10.1103/PhysRevLett.110.096101}

\bibitem{ZhenHann19}
C.~X. Zheng, K.~Hannikainen, Y.~R. Niu, J.~Tersoff, D.~Gomez, J.~Pereiro, D.~E.
  Jesson,
  \href{https://link.aps.org/doi/10.1103/PhysRevMaterials.3.124603}{Mapping the
  surface phase diagram of gaas(001) using droplet epitaxy}, Phys. Rev.
  Materials {\bf{3}} (2019) 124603.
\newblock \href {http://dx.doi.org/10.1103/PhysRevMaterials.3.124603}
  {\path{doi:10.1103/PhysRevMaterials.3.124603}}.
\newline\urlprefix\url{https://link.aps.org/doi/10.1103/PhysRevMaterials.3.124603}

\bibitem{AquaGouy11}
J.-N. Aqua, A.~Gouy\'e, T.~Auphan, T.~Frisch, A.~Ronda, I.~Berbezier,
  \href{http://dx.doi.org/10.1063/1.3576916}{Orientation dependence of the
  elastic instability on strained {SiGe} films}, Appl. Phys. Lett. 98 (2011)
  161909.
\newblock \href {http://dx.doi.org/10.1103/PhysRevB.82.085322}
  {\path{doi:10.1103/PhysRevB.82.085322}}.
\newline\urlprefix\url{http://dx.doi.org/10.1063/1.3576916}

\end{thebibliography}

\end{document}